\documentclass{ws-procs9x6}

\begin{document}

\title{$J/\psi$ Photo-production at Large $z$ in Soft Collinear Effective Theory}

\author{Sean Fleming\footnote{Presentation given at the Ringberg Workshop: New Trends in HERA Physics 2005.}}

\address{Physics Department \\
University of Arizona, \\ 
Tucson, AZ 85721, USA\\ 
E-mail: fleming@physics.arizona.edu}

\author{Adam K. Leibovich}

\address{Department of Physics and Astronomy \\ 
University of Pittsburgh \\
Pittsburgh, PA 15260, USA\\
E-mail: akl2@pitt.edu}  

\author{Thomas Mehen}

\address{Department of Physics \\
Duke University \\
Durham, NC 27708, USA \\
and \\
Jefferson Laboratory \\
12000 Jefferson Ave. \\
Newport News, VA 23606. USA \\
E-mail: mehen@phy.duke.edu} 

\maketitle

\abstracts{One of the outstanding problems in $J/\psi$ physics is a systematic understanding of the differential photo-production cross section $d\sigma/dz (\gamma + p \to J/\psi + X$), where $z= E_\psi/E_\gamma$ in the proton rest frame. The theoretical prediction based on the non-relativistic QCD (NRQCD) factorization formalism has a color-octet contribution which grows rapidly in the endpoint region, $z \to 1$, spoiling perturbation theory. In addition there are subleading operators which are enhanced by powers of $1/(1-z)$ and they must be resummed to all orders. Here an update of a systematic analysis is presented. The approach used to organize the endpoint behavior of the photo-production cross section is based on a combination of NRQCD and soft collinear effective theory. While a final result is not yet available, an intermediate result indicates that better agreement between theory and data will be achieved in this framework.
}

\section{Introduction}

The production of $J/\psi$ in high energy electron-proton interactions has proven to be a rich subject with great potential for furthering our understanding of the perturbative and non-perturbative regimes of QCD. In particular a number of different kinematic regions are accessible in $ep$ collisions so that a range of different $J/\psi$ production phenomena can be explored with a single experiment. The H1 and Zeus collaborations have gathered a large amount of data on $J/\psi$ production and study a diverse set of topics including processes where the scattered electron is in the forward region so that the photon exchanged with the proton is real.

The theoretical framework within which the production of non-relativistic bound states can be systematically treated is the NRQCD factorization formalism~\cite{bbl}. This approach relies on a non-relativistic effective theory of QCD~\cite{bbl,Brambilla:1999xf,lmr}, which is a systematic expansion of QCD about the limit of vanishing quark velocity: $v \to 0$. The resulting effective field theory consists of an infinite number of operators each scaling as a definite power of $v$, however, at at a given order in $v$ there are only a finite number of terms. As a consequence working to a specified numerical accuracy only requires including those contributions that affect the calculation at the level of significance desired. 

Reference~[\refcite{bbl}] postulates that the $J/\psi$ photo-production cross section in $ep$ collisions is of the form
\begin{equation}
\label{NRQCDFF}
d\sigma(\gamma+p \to J/\psi + X) = f_{i/p} \otimes d \hat{\sigma}(i+\gamma \to  c\bar{c}[n] + \hat{X}) \otimes F(c\bar{c}[n] \to J/\psi) \,,
\end{equation}
where $d \hat{\sigma}(i+\gamma \to c\bar{c}[n] + \hat{X})$ is the short-distance scattering cross section for producing a $c\bar{c}$ pair with non-relativistic relative momentum in a color and angular momentum state indexed by $n$. The parton distribution function  $f_{i/p}$ gives the probability for finding parton $i$ in the proton, and $F(c\bar{c}[n] \to J/\psi)$ gives the probability for the $c\bar{c}$ pair with quantum numbers $n$ to hadronize into a $J/\psi$. There is an implied sum over quantum numbers $n$, with each $F(c\bar{c}[n] \to J/\psi)$ scaling as $v^{\gamma(n)}$, with $\gamma(n)$ the scaling dimension of the operator. The short-distance coefficient $d \hat{\sigma}(i+\gamma \to c\bar{c}[n] + \hat{X})$ scales with powers of the strong coupling constant $\alpha_s$, and can be enhanced by kinematic factors as well. As a result there is a competition between suppression by $v^{\gamma(n)}$ and enhancements of $d\hat{\sigma}$, and extra care must be taken when deciding which terms to keep in Eq.~(\ref{NRQCDFF}). In particular there are two effects that can greatly enhance the short-distance cross section: fragmentation contributions~\cite{Braaten:1993mp,Braaten:1993rw} and color-octet contributions~\cite{spf6}.

\begin{figure}[ht]
\centerline{\epsfxsize=8cm \epsfbox{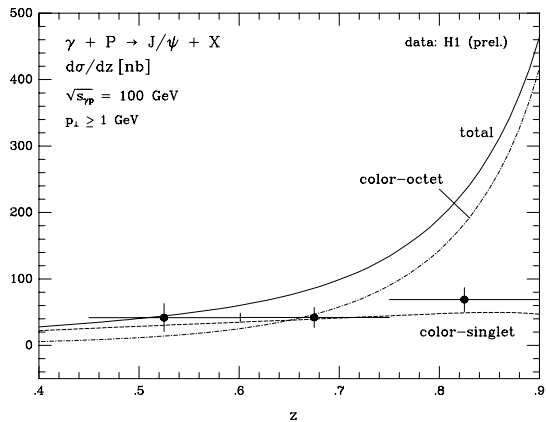}}
\caption{The $J/\psi$ energy distribution $d\sigma/dz$ at the photon-proton center-of-mass energy $\sqrt{s} = 100$ GeV integrated in the range $p_\perp \geq 1$ GeV from Ref.~[7].}
\label{photopro1}
\end{figure}

In Refs.~[\refcite{Cacciari:1996dg,Amundson:1996ik,Ko:1996xw}] the total $J/\psi$ photo-production cross section $\sigma(\gamma + p \to J/\psi +X)$ is calculated and compared to HERA data as well as other experimental data. The conclusion is that theory and data agree as long as the non-perturbative NRQCD production matrix elements are allowed to be negative. However, given the interpretation of the NRQCD matrix elements as the probability for a $c\bar{c}$ pair to fragment into a $J/\psi$ and any number of soft particles there is a prejudice against negative values of the matrix elements. Furthermore there is uncertainty whether the NRQCD factorization formalism is valid for the total cross section since the $J/\psi$ is not necessarily produced with large transverse momentum. References~[\refcite{Cacciari:1996dg,Ko:1996xw}] also considered the differential cross section $d\sigma/ d z (\gamma + p \to J/\psi(z) +X)$ where $z = p_p\cdot p_\psi / p_p \cdot p_\gamma$ for $J/\psi$ transverse momentum greater than 1 GeV.
The results of Ref.~[\refcite{Cacciari:1996dg}] compared to H1 data~\cite{H1:1} are shown in Fig.~\ref{photopro1}.
The theoretical curve is an increasing function of $z$ while the experimental points indicate a flat distribution, and it is clear that the color-octet contribution dominates the differential cross section at larger values of $z$. The rise in the cross section is due to terms of the form
\begin{equation}
\left( \frac{\ln(1-z)}{1-z}\right)_+ \,, \qquad  
\left( \frac{1}{1-z}\right)_+ \,,
\end{equation}
which arise at next-to-leading order in the color-octet differential cross section. These plus-distributions become arbitrarily large as $z \to 1$, and cause a breakdown of perturbation theory. As a consequence such terms must be resummed to make a sensible comparison of theory to data in the large $z$ region.

\begin{figure}[ht]
\centerline{\epsfxsize=9cm \epsfbox{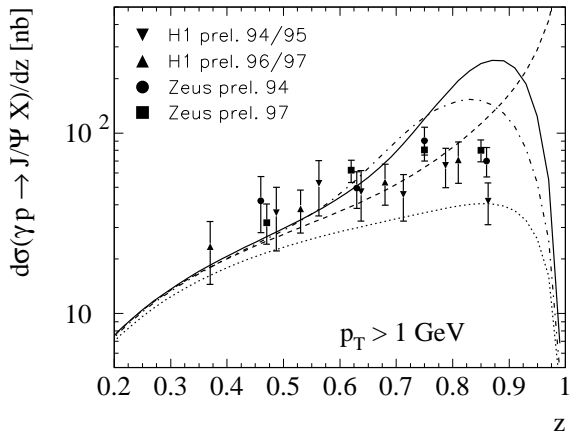}}
\caption{The $J/\psi$ energy spectrum at $\sqrt{s} = 100$ GeV integrated in the range $p_\perp \geq 1$ GeV compared to HERA data. Solid, dashed-dotted, and dashed lines are the color-singlet contribution plus the color-octet shape function contribution for three different values of a shape function parameter. The dotted line is the color-singlet contribution alone. Plot from Ref.~[13].}
\label{shapefunc}
\end{figure}

In addition to the breakdown of perturbation theory, the non-perturbative NRQCD expansion breaks down when $z \to 1$, because of a set of subleading operators which scale as $v^{2n}/(1-z)^n$. This was first explored in Refs.~[\refcite{Rothstein:1997ac,Beneke:1997qw}], where the infinite tower of enhanced operators was resummed into a shape function. In Ref.~[\refcite{Beneke:1999gq}] an analysis of the differential cross section $d\sigma/dz$ including the shape function was carried out. The spectrum compared to H1 and Zeus data~\cite{HERA} is shown in Fig.~\ref{shapefunc}. 
The introduction of a shape function tames the endpoint divergences of the color-octet contribution at NLO, however, agreement with data is not great. In particular the shoulder in the theoretical prediction is too large and peaked too far to the right. This is due to the inconsistent treatment of the plus-distribution terms which also need to be resummed before a comparison of theory and data can be made. In Ref.~[\refcite{Fleming:2003gt}] it was shown that in $e^+e^- \to J/\psi + X$ at the endpoint perturbative resummation in addition to the shape function significantly broadens the differential distribution relative to either including only the shape function or only carrying out a perturbative resummation without shape function. 

\section{The breakdown of the NRQCD expansion}

It is helpful to study the kinematics of $\gamma + p \to J/\psi +X$ as $z \to 1$ to gain a better understanding of the physics that leads to the breakdown of NRQCD factorization in this kinematic regime. In the proton-photon center-of-mass frame
\begin{equation}
p_\gamma^\mu = \frac12\sqrt{s} \bar n^\mu, \qquad 
p_p^\mu = \frac12\sqrt{s} n^\mu, \qquad
p_{c\bar{c}}^\mu = M v^\mu + k^\mu \,,
\label{MomDef}
\end{equation}
where $n^\mu = (1,0,0,-1)$, $\bar{n}^\mu = (1,0,0,1)$, $M = 2m_c$, and $v^\mu$ and $k^\mu$ are the 4-velocity of the $J/\psi$ and the residual momentum of the $c\bar c$ pair in the $J/\psi$ respectively. 
In terms of the scaling variable $z$ the $J/\psi$ velocity is
\begin{equation}
p_\psi^\mu = M_\psi v^\mu = \frac{z\sqrt{s}}{2} \bar n^\mu + p_\perp^\mu + \frac{M_\perp^2}{2 z \sqrt{s}}n^\mu
\end{equation}
where $M_\perp^2 = M_\psi^2 - p_\perp^2$, with $p^2_\perp = -\textbf{p}^2_\perp$.
By momentum conservation 
\begin{eqnarray}
p_X^\mu &=& p_\gamma^\mu + p_p^\mu - p_{c\bar{c}}^\mu 
\nonumber \\
&=& \frac{1}{2} \frac{\sqrt{s}M}{M_\psi} \bigg( \frac{M_\psi}{M} -  z \bigg) \bar{n}^\mu 
+ \frac{1}{2} \sqrt{s} \bigg( 1 -  \frac{MM^2_\perp}{sz M_\psi } \bigg) n^\mu
- \frac{M}{M_\psi}p^\mu_\perp-k^\mu  \,.
\end{eqnarray}
Setting $k = 0$ the invariant mass squared of the final state is
\begin{equation}
p^2_X =  \frac{sM}{M_\psi} \bigg( \frac{M_\psi}{M} -  z \bigg)\bigg( 1 -  \frac{MM^2_\perp}{sz M_\psi } \bigg) 
+  \frac{M^2}{M_\psi^2} p^2_\perp \,.
\label{XMassSqr}
\end{equation}
This is much larger than $\Lambda\sim 1 \,, \textrm{GeV}$ as long as $ M_\psi/M - z \sim 1$. Note the requirement for $p^2_X \geq 0$ translates into $\textbf{p}^2_\perp \leq (s z - MM_\psi)(M_\psi/M - z)$ so $p_\perp$ is forced to be small at the endpoint. Thus away from the endpoint the decay products can be integrated out and NRQCD factorization holds. However if $z \to M_\psi/M$ the invariant mass of the decay products becomes much smaller than the $\bar{n}\!\cdot\!p_X$ momentum component and the final state consists of a $J/\psi$ and a jet in the direction of the incoming hadron. In this kinematic regime NRQCD factorization no longer holds, and an effective field theory which describes the light-like final state must be used. The appropriate theoretical framework is a combination of NRQCD to describe the $J/\psi$ and soft collinear effective theory (SCET)~\cite{Bauer:2001ew,Bauer:2001yr,Bauer:2001ct,Bauer:2001yt} to describe the light-like decay products. A combination of SCET and NRQCD has already been used to study a variety of Quarkonium decay and production processes in kinematically restricted regions~\cite{Bauer:2001rh,Fleming:2002rv,Fleming:2002sr,Fleming:2003gt,GarciaiTormo:2004jw,Fleming:2004rk,Fleming:2004hc,GarciaiTormo:2005ch}.

Before diving into a detailed description of SCET it is useful to determine the region in $z$ where SCET is applicable. To do this return to Eq.~(\ref{XMassSqr}) and include the residual momentum of the $c\bar{c}$ pair introduced in Eq.~(\ref{MomDef}). The scaling of the residual momentum $k^\mu$ can be found by boosting from the $J/\psi$ rest frame where $k^\mu \sim \Lambda$ to the proton-photon center of mass frame. In light-cone coordinates
\begin{equation}
k^\mu \sim 
\Lambda \left( \frac{ \sqrt{s}}{M_\psi}, \frac{M_\psi}{\sqrt{s}}, 1\right) \,,
\end{equation}
where $z \sim 1$ has been used. Note the plus light-cone component is enhanced by $\sqrt{s}/M_\psi$, while the minus component is suppressed by the same amount. Including the residual momentum in Eq.~(\ref{XMassSqr}) gives
\begin{equation}
p^2_X =  \frac{sM}{M_\psi} \bigg( \frac{M_\psi}{M} -  z \bigg)\bigg( 1 -  \frac{MM^2_\perp}{sz M_\psi } \bigg) 
+ \frac{M^2}{M_\psi^2} p_\perp^2- \sqrt{s}  \bigg( 1 -  \frac{MM^2_\perp}{sz M_\psi } \bigg) n \cdot k + ...
\end{equation}
where suppressed pieces are dropped. The last term in the equation above scales as $s \Lambda/ M_\psi$ so the final state becomes sensitive to nonperturbative physics when $z \gtrsim (M_\psi - \Lambda)/M$. 

\section{Soft Collinear Effective Theory}

SCET describes the dynamics of highly energetic particles moving close to the light-cone interacting with a background field of soft quanta. The interaction of the collinear particle with the background introduces a small residual momentum component into the light-like collinear momentum so that collinear particles have momentum $p^\mu = Q n^\mu+k^\mu$,  where $k^\mu \sim \Lambda$. However, collinear particle do not only interact with the soft background, they also couple to other collinear particles. As a consequence the SCET Lagrangian consists of two sectors: soft and collinear. The SCET Lagrangian was first derived in what is called the label formalism~\cite{Bauer:2001ew,Bauer:2001yr,Bauer:2001ct,Bauer:2001yt}, and was subsequently formulated in position space~\cite{Beneke:1,Beneke:2}. To illustrate some important properties of SCET consider the quark sector of the Lagrangian. It can be split into two pieces: one  coupling collinear to soft
\begin{equation}\label{scetlag}
\mathcal{L}_{cs} = \bar{\xi}_{n,p}  i n\cdot D  \frac{{\bar n\!\!\!\slash}}{2} \xi_{n,p} \,,
\end{equation}
where $\xi_{n,p}$ is the collinear quark field labelled by collinear momentum $p^\mu = n^\mu \, \bar{n}\cdot p /2  + p^\mu_\perp$, $\bar{n} = (1,0,0,1)$, and $i n\cdot D = i n \cdot \partial +  g n\cdot A_{s}$, with $A_{s}$ the soft gauge field. This expression looks very much like the HQET Lagrangian with the velocity $v^\mu$ replaced with the light-like vector $n^\mu$, and was first derived in Ref.~[\refcite{Dugan:1990de}]. The second piece of the collinear Lagrangian consists of interactions of only collinear particles among themselves
\begin{equation}
\mathcal{L}_{c} = \bar{\xi}_{n,p'} \bigg\{  g n\cdot A_{n, q} + i {D\!\!\!\!\slash}^\perp_c
\frac{1}{i\bar{n} \cdot D_c} i   {D\!\!\!\!\slash}^\perp_c \bigg\} \frac{{\bar n\!\!\!\slash}}{2}
\xi_{n,p}
\,,
\end{equation}
where $A_{n, q}$ is the collinear gauge field, and the collinear covariant derivative is $i D^\mu_c = \mathcal{P}^\mu + ig A_{n,q}$. The operator $\mathcal{P}^\mu$ projects out collinear label momentum: $\mathcal{P}^\mu \xi_{n,p} = (n^\mu  \, \bar{n}\cdot p /2 + p^\mu_\perp) \xi_{n,p}$~\cite{Bauer:2001yr}. 
The SCET Lagrangian is invariant under separate collinear and soft gauge transformations which provide a powerful restriction on the operators allowed in the theory~\cite{Bauer:2001yt}. Furthermore the Lagrangian is invariant under a global $U(1)$ helicity spin symmetry, and must be invariant under certain types of reparameterizations of the collinear sector of the Lagrangian~\cite{Manohar:2002fd,Chay:2002mw}. 

A remarkable consequence of the gauge symmetries of SCET is the factorization of soft and collinear effects. Towards this end the soft Wilson line is introduced
\begin{equation}
Y(x) = {\rm P exp} \bigg( ig \int^x_{-\infty} ds \; n\cdot A_{us}(ns) \bigg) \,,
\end{equation}
and the collinear fields are redefined as follows:
\begin{equation}\label{fieldred}
\xi_{n,p} = Y \xi^{(0)}_{n,p}   \hspace{5ex} A^\mu_{n,q} = Y A^{(0) \mu}_{n,q} Y^\dagger \,.
\end{equation}
After the field redefinitions the soft gluons decouple from the collinear fields, {\it i.e.} $\mathcal{L}_{cs} \to 0$ in Eq.~(\ref{scetlag}), and the collinear Lagrangian becomes independent of soft physics. At higher orders in the SCET expansion this decoupling does not occur, and factorization is broken. 

\section{Factorization at the Endpoint}

The soft-collinear factorization properties of SCET can be used to obtain a factored form for the $J/\psi$ production cross section at the kinematic endpoint. An important condition for the derivation to be valid is that a sufficient range of $p_\perp$ is integrated over. To be precise we must smear over a range $p_\perp \gtrsim \Lambda M$. If this is not done, or if only a range $p_\perp \sim \Lambda$ is integrated over the intrinsic transverse momentum of the partons in the proton must be taken into account~\cite{Sridhar:1998rt,Baranov:2003at,Lipatov:2002tc}. This is an important point because an experimental cut of $p_\perp \geq 1 \,,\textrm{GeV}$ is usually implemented in HERA data. This cut lies in the regime where non-perturbative momentum is important and may introduce a sensitivity to the transverse momentum of the partons in the proton.

There are two steps in factorizing the photo-production differential cross section at the endpoint. The first step is matching QCD onto SCET where the collinear particles have a typical off-shellness of order $\Lambda M$. This formulation of SCET is called SCET${}_{\rm I}$. In the next step SCET${}_{\rm I}$ is matched onto SCET${}_{\rm II}$ where collinear fields have a typical off-shellness $\mathcal{O}(\Lambda^2)$~\cite{Bauer:2002aj}. To be concrete one of the two leading color-octet contributions to $J/\psi$ photo-production at the endpoint is considered: a $c\bar{c}$ pair produced in a color-octet ${}^1S_0$ configuration which hadronizes to $J/\psi$ via a chromomagnetic spin-flip transition.

The derivation is easiest in the parton-photon CM frame where
\begin{equation}
p_\gamma^\mu = \frac12\sqrt{\hat s} \bar n^\mu, 
p_g = x p_P^\mu = \frac12  \sqrt{\hat s} n^\mu,  
p_\psi^\mu =   \frac{z\sqrt{\hat s}}{2} \bar n^\mu + p_\perp^\mu + \frac{M_\perp^2}{2 z \sqrt{\hat s}}n^\mu \,,
\end{equation}
with the $J/\psi$ approximately at rest, $v^\mu =(1,0,0,0)$. First the QCD current is matched onto the SCET${}_{\rm I}$ current
\begin{eqnarray}\label{currmatch}
J^\mu(x) &=&\sum_{\mbox{\boldmath $\omega$}} \,
e^{i(Mv - \mbox{\boldmath $\omega$} )\cdot x} C^{\mu}_{ \alpha}(\omega) 
J_{\mbox{\boldmath $\omega$}}^\alpha( x) \,,
\nonumber \\
J_{\mbox{\boldmath $\omega$}}^\alpha (x) &=& 
\big[ \psi^\dagger_{{\bf p}} \delta^{(3)}_{\vec{\mathcal{P}} , \mbox{\boldmath $\omega$}}B^\alpha_{\perp} \chi_{-{\bf p}} \big] (x) \,,
\end{eqnarray}
where $\mbox{\boldmath $\omega$} = \omega n^\mu/2 + \omega^\mu_\perp$, and $\vec{\mathcal{P}} = \mathcal{P} n^\mu/2 + \mathcal{P}_\perp$. At lowest order in $\alpha_s(M)$ 
\begin{eqnarray}\label{match}
C_{\mu_\alpha} =\frac{2 e e_c g_s}{M} \epsilon^\perp_{\mu \alpha} \,.
\end{eqnarray} 

The second step is to integrate out the scale $\Lambda M$ by match the differential cross section in SCET${}_{\rm I}$ onto SCET${}_{\rm II}$. To derive the cross section it is best to take a step back and consider the differential cross section in QCD:
\begin{equation}
2 E_\psi \frac{d\sigma}{d^3 p_\psi} = 
\frac{-g^{\mu \nu}}{16 \pi^3 s}  \sum_{X} \int d^4 y \, e^{-i p_\gamma \cdot y}
\langle p | J^\dagger_\mu(0) | J/\psi + X \rangle 
\langle J/\psi + X | J_\nu(y) | p \rangle 
\end{equation}
Inserting the expression in Eq.~(\ref{currmatch}) gives the SCET${}_{\rm I}$ cross section 
\begin{eqnarray}
2 E_\psi \frac{d\sigma}{d^3 p_\psi} &=&\frac{-g_{\mu \nu}}{16 \pi^3 s} \sum_{\mbox{\boldmath $\omega$}_{1,2}}
[C^\mu_\alpha({\mbox{\boldmath $\omega$}}_1)]^\dagger  C^\nu_\beta({\mbox{\boldmath $\omega$}}_2)  
 \int d^4 y \, e^{-i (p_\gamma -M v +\mbox{\boldmath $\omega$}_2)\cdot y} \nonumber \\
& & \hspace{-15ex}
\times \langle p | [\chi^\dagger_{{\bf -p}} \delta^{(3)}_{\vec{\mathcal{P}} , \mbox{\boldmath $\omega$}}B^\alpha_{\perp} \psi_{{\bf p}}](0)| J/\psi + X \rangle 
\langle J/\psi + X |  [ \psi^\dagger_{{\bf p}} \delta^{(3)}_{\vec{\mathcal{P}} , \mbox{\boldmath $\omega$}}B^\alpha_{\perp} \chi_{-{\bf p}}](y) | p \rangle \,,
\end{eqnarray}
and soft-collinear factorization properties of SCET outlined in the previous section can be used to obtain a factored form for this cross section:
\begin{equation} \label{dsdz}
\frac{d\sigma}{dz} =
\frac{\alpha \alpha_s e_c^2}{\pi M z s}   \langle\mathcal{O}^{\psi}_8(^1S_0)  \rangle
\int d n\cdot k \, \hat S\bigg(\sqrt{\hat s}\left(1- \frac{M}{M_\psi} z\right) + n\cdot k\bigg) \mathcal{J}_{\mbox{\boldmath $\omega$}_+}(n\!\cdot\! k) \,.
\end{equation}
Here the momentum space functions $\mathcal{J}$ and  $\hat S$ are Fourier transforms of collinear and soft matrix elements respectively:
\begin{eqnarray}\label{two}
\langle p_I |  \textrm{Tr}\big[  B_\perp^{(0)}(0)  \delta_{\vec{\mathcal{P}}_+ , \mbox{\boldmath $\omega$}_+}
 B_\perp^{(0)}(y) \big] | p_I \rangle  && \nonumber \\
&& \hspace{-20ex}
 = \delta(n\!\cdot\! y) \delta^{(2)}(y_\perp) \int \frac{d n\!\cdot\! k}{2 \pi}
e^{-\frac{i}{2}n\cdot k \bar{n}\cdot y} \mathcal{J}_{\mbox{\boldmath $\omega$}_+}(n\!\cdot\! k) \,.  
\end{eqnarray}
and
\begin{eqnarray}\label{sdef}
 S_\psi(\omega) &\equiv& 2M \langle\mathcal{O}^{\psi}_8(^1S_0)\rangle 
 \hat S_\psi(\omega)  \nonumber \\
& & \hspace{-10ex} = 
 \int \frac{d \bar{n}\!\cdot\! y}{4 \pi} e^{-\frac{i}{2}\omega \bar{n}\cdot y}
 \langle 0 | \chi^\dagger_{-{\bf p}} Y T^A Y^\dagger \psi_{{\bf p}} (0) \mathcal{P}_\psi 
 \psi^\dagger_{{\bf p}'} Y T^A Y^\dagger \chi_{-{\bf p}'}(\bar{n}\!\cdot\! y) |0\rangle \,.
\end{eqnarray}
By construction $\int d\omega \hat S(\omega) = 1$.  

Next SCET${}_{\rm I}$ is matched onto SCET${}_{\rm II}$. This entails performing an operator product expansion (OPE) on $\mathcal{J}_{\mbox{\boldmath $\omega$}_+}(n\!\cdot\! k)$, where the large scale is set by the invariant mass of the collinear degrees of freedom in SCET${}_{\rm I}$: 
\begin{eqnarray}\label{scet2match}
\mathcal{J}_{\mbox{\boldmath $\omega$}}(n\!\cdot\! k) & \approx &  
C_{II}\bigg( \frac{n\!\cdot\! k \, }{\omega_+}\bigg) \delta^{(2)}_{0,\omega^\perp_+}
\nonumber \\
& & \times
\langle p_{II} | 
\bigg[  \textrm{Tr}\big\{ T^B B_\perp^{II}(0) \big\}  \delta_{\bar{n}P_+ ,\omega_+} 
 \textrm{Tr}\big\{ T^B B_\perp^{II}(0) \big\}  \bigg]
| p_{II} \rangle \,,
\end{eqnarray}
where corrections are suppressed by $\mathcal{O}(\Lambda/M)$. 
The coefficient function is dimensionless, and therefore must be a function of the ratio of the invariant mass squared and the large momentum component squared. Furthermore, at leading order in $\alpha_s(\Lambda M)$ the perpendicular momentum of the $J/\psi$ is $\mathcal{O}(\Lambda)$ so that the labels $\omega_+^\perp$ must be zero. The SCET${}_{\rm II}$ operator above can be related to the familiar parton distribution function which gives the probability of finding a gluon in the proton~\cite{Bauer:2002nz}
\begin{eqnarray}
\frac{1}{2} \sum_{\textrm{spin}} 
\langle p_{II} | 
\bigg[  \textrm{Tr}\big\{ T^B B_\perp^{II}(0) \big\}  \delta_{\bar{n}P_+ ,\omega_+} 
 \textrm{Tr}\big\{ T^B B_\perp^{II}(0) \big\}  \bigg]
| p_{II} \rangle && \nonumber \\
&& \hspace{-40 ex}
= -n\cdot\! p_p \int^1_0 dx \, \delta(\omega_+ - 2 x n\cdot\! P_p ) f_{g/p}(x) \,,
\end{eqnarray}
where $f_{g/p}(x)$ is the parton distribution function. Using this result the final factored form of the cross section is
\begin{equation}\label{cs}
\frac{d \sigma}{d z} = M \big| C^{\mu\alpha}(M) \big|^2
\sum_{v_\perp} \delta^{(2)}_{0,v_\perp}
\int d \xi  \, S_\psi(\xi) \, C_{II}( \xi -1+z - \overline{\Lambda})
\, f_{g/p}\bigg( \frac{M^2}{s}\bigg)\,.
\end{equation}

\section{Summing Logarithms}

\begin{figure}
\begin{center}
\includegraphics[width=10cm]{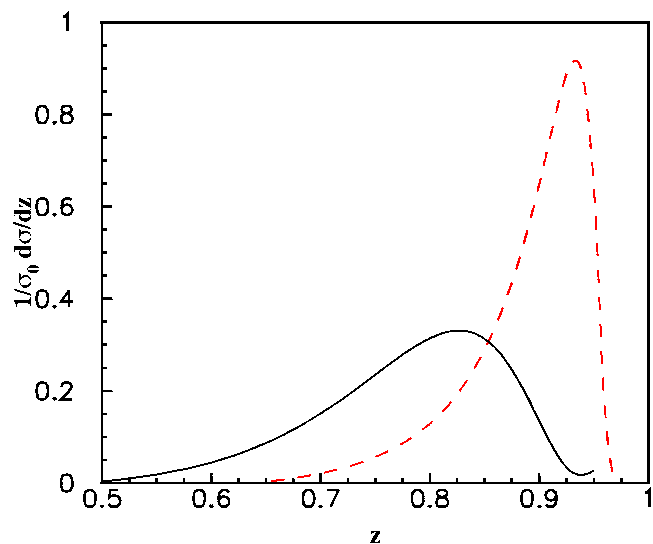}
\caption{The differential cross section in units of the total color-octet ${}^1S_0$ cross section. The dashed curve only includes resummation of singular plus-distribution terms arising in perturbation theory. The solid line includes both non-perturbative resummation in the form of the shape function and perturbative resummation. }
\label{plot}
\end{center}
\end{figure}

Large logarithms are summed using the renormalization group equations (RGEs), which is a two step procedure in this case. First the effective theory currents in Eq.~(\ref{currmatch}) are run from the scale $\mu_H = \sqrt{s}$ to the scale $\mu_{\textrm{I}} = M\sqrt{1-z}$ using the SCET${}_{\textrm{I}}$ RGEs. Next the SCET${}_{\textrm{II}}$ operators in Eq.~(\ref{cs}) are run to the scale $\mu\sim\Lambda$. 
Details of the running are left for a later publication~\cite{inprog}. Here preliminary results are presented in Fig.~\ref{plot}, which shows the differential cross section (in units of the total color-octet ${}^1S_0$ photo-production cross section). The solid line includes both perturbative resummation and non-perturbative resummation {\it i.e.} the shape function. The dashed line includes only perturbative resummation. 
This should be compared to Fig.~\ref{shapefunc} where only the shape function is included. In the case where either the shape function alone is used, or only perturbation theory is resummed the spectrum is too sharply peaked to be compatible with data. However, when both the shape function and perturbative resummation is included the spectrum softens considerably, which gives hope that theory will be compatible with data in a complete analysis. 

\section*{Acknowledgments}
A.L.~was supported in part by the National Science 
Foundation under Grant No.~PHY-0244599. T.M.~was supported in part by the Department
of Energy under grant numbers DE-FG02-96ER40945 and DE-AC05-84ER40150.

%%%%%%%%%%%%%%%%%%%%%%%%%%%%%%%%%%%%%%%%%%
%Bibliography

\end{document}